# From Full-fledged ERP Systems Towards Process-centric Business Process Platforms

*Completed Research Full Paper*

**Lukas Böhme**
HPI, University of Potsdam
lukas.boehme@hpi.de

**Tobias Wuttke**
HPI, University of Potsdam
tobias.wuttke@hpi.de

**Benedict Bender**
LSWI, University of Potsdam
benedict.bender@wi.uni-potsdam.de

**Ralf Teusner**
HPI, University of Potsdam
ralf.teusner@hpi.de

**Sebastian Baltes**
SAP SE
sebastian.baltes@sap.com

**Christoph Matthies**
HPI, University of Potsdam
christoph.matthies@hpi.de

**Michael Perscheid**
HPI, University of Potsdam
michael.perscheid@hpi.de

## Abstract

Enterprise Resource Planning (ERP) systems are critical to the success of enterprises, facilitating business operations through standardized digital processes. However, existing ERP systems are unsuitable for startups and small and medium-sized enterprises that grow quickly and require adaptable solutions with low barriers to entry. Drawing upon 15 explorative interviews with industry experts, we examine the challenges of current ERP systems using the task technology fit theory across companies of varying sizes. We describe high entry barriers, high costs of implementing implicit processes, and insufficient interoperability of already employed tools. We present a vision of a future business process platform based on three enablers: Business processes as first-class entities, semantic data and processes, and cloud-native elasticity and high availability. We discuss how these enablers address current ERP systems' challenges and how they may be used for research on the next generation of business software for tomorrow's enterprises.

## Introduction

Enterprise resource planning (ERP) systems play a vital role in the application landscape in most of today's enterprises. They offer comprehensive standard solutions and facilitate value-adding business processes for enterprises worldwide (Asprion et al., 2018). However, in practice, we observe several unsolved challenges of ERP systems, such as limited process adaptability and tedious upgrade projects (Abd Elmonem et al., 2016; Bender et al., 2021; Sancar Gozukara et al., 2022). Especially startups and small and medium-sized enterprises (SMEs) defer implementing ERP systems, according to a survey by the German Federal Statistical Office (Destatis, 2021). But even if ERP systems are implemented, it is not guaranteed that the system will operate unchanged for an extended period. Changing market requirements force companies to adapt their ERP system implementations, resulting in migration projects and complex process adjustments (Bender et al., 2021; Yusuf et al., 1999).

The identification and understanding of challenges related to ERP systems are the focus of current ERP research (Abd Elmonem et al., 2016; Bender et al., 2021; Sancar Gozukara et al., 2022). However, companies' requirements are constantly changing (Hvolby & Trienekens, 2010) and, thus, the business processes they support. Therefore, we gathered first-hand insights into the challenges of state-of-the-art ERP systems by conducting 15 semi-structured interviews with industry experts working in a diverse set of companies ranging from startups to international corporations and covering various domains, including manufacturing, software development, and e-learning.

Through the theoretical lens of task-technology fit (TTF) and the use of a Gioia Matrix, we identified three main challenges of current ERP systems for growing companies. First, ERP processes are implicit and offer insufficient transparency allowing only a small group to understand the underlying business logic. Second, ERP systems have a high entry barrier due to their complexity and costly implementation. Especially for startups, ERP systems offer vast components that some perceive as irrelevant to their current needs or hinder the company's development due to their non-transparent and inter-dependent procedures. Finally, ERP systems pose integration challenges since business processes span multiple IT systems and might even cross company boundaries.

To address the identified challenges, this paper presents a vision for future enterprise systems to overcome these challenges: process-centric *business process platforms* (BPPs). Our focus lies on examining the high-level concepts that enable such platforms. Hence, we pose the following research question: *What are enabling technologies a business process platform requires to seamlessly support companies' business processes while growing?* We suggest that BPPs employ adaptable, executable business process models, semantically-enriched data, and cloud-native development techniques. Based on our findings, we present our vision of a BPP founded on core services, visually modeled, executable business processes, and a semantic domain model.

## Theoretical Background

This section provides a brief overview of business process management (BPM), the adaptability of enterprise systems, and the task-technology fit theory. Thus, we lay the foundation to understand the identified challenges of ERP systems.

BPM is a management discipline structuring business processes to improve organizations' performance (Vukšić et al., 2017). The foundation of BPM is the recognition that the production of each product offered by a company to the market results from a series of different activities. These activities are organized and improved using business processes. Enterprise systems, such as ERP systems, play a crucial role in BPM, as they support many of the tasks performed by a company. These tasks can be enacted manually by employees or through information systems (Weske, 2012).

The ability to adapt to changing conditions is considered a key success factor for organizations (Da Xu, 2011). Environmental (external) changes or internal changes such as rapid growth require companies to adopt. Given the wide implementation of enterprise systems, the adaptability of companies is determined by the adaptability of its enterprise systems. Prior studies consider adaptable enterprise systems as a critical requirement of companies' survival (Almutairi et al., 2022). The need to adopt enterprise systems to individual circumstances is widely acknowledged (Davenport, 1998). In this regard, different possibilities to conduct changes in enterprise systems evolved (Haines, 2009). With the increased importance of process organization, ERP system flexibility is determined by their possibility to adapt implemented processes. Current challenges and potential drivers for tomorrow's enterprises are suggested to support process adoption and company growth.

The modularity of systems to enact business processes was already the subject of research. Pauker et al. (2018), for instance, introduce centurio.work, a modular manufacturing execution system based on a service-oriented oriented approach and Business Process Model and Notation (BPMN). Another approach is presented by Künzle and Reichert (2011), who describe a framework called PHILharmonicFlows that operates on an object-based approach to integrate data, processes, and users. While these existing approaches focus on rigorously following an object-aware paradigm or industry-specific use cases, we aim to provide first steps toward Gartner's vision of the "composable enterprise" (Gartner, 2020).

Task-technology fit (TTF) is a theory that examines the compatibility between a technology and a particular task. Developed initially by Goodhue and Thompson (1995), TTF was intended to be applied at the individual level. However, Zigurs and Buckland (1998) modified the theory to apply to groups as well. The TTF theory is widely accepted in the Information Systems research community, and it is frequently used to analyze various contexts (e.g., (Lim and Benbasat, 2000) or (Ferratt and Vlahos, 1998)).

In this study, we evaluate the utilization of ERP systems by growing companies using TTF. As companies expand, their organizational structure and tasks become more complex, and they may require new technological solutions to manage their business processes effectively. ERP systems are designed to integrate all aspects of a company's operations into a single software system. However, the success of ERP implementation depends on various factors, including the match of the technology with the company's tasks and requirements. Thus, by utilizing the TTF theory, we analyze how well ERP systems meet the specific needs of growing companies. We examine whether the ERP systems used by these companies are suitable for their tasks and organizational structure and identify areas where improvements can be made.

## Methodology

The development and usage of business applications is practice-oriented and rapidly evolving. Given this inherent nature, we conducted an exploratory qualitative research study by interviewing practitioners to gain an up-to-date perspective on the prevailing problems of today's ERP and business systems. We decided for purposeful sampling due to the limited number of available interviewees (Palinkas et al., 2015). Overall, we conducted 15 semi-structured interviews with industry experts from companies operating in technology, manufacturing, e-commerce, and logistics located in German-speaking countries in the timespan between December 2021 and the end of April 2022. The interviewees worked in companies ranging from recently founded startups to established SMEs with up to 200 employees or large multinational enterprises. The interviewees either had the position of a founder or a leading role in the technical implementation of the company's business or ERP systems. They were selected from the authors' network based on their ERP and business software expertise. Furthermore, we validated our findings with six ERP system experts from an established ERP system vendor.

Each interview was structured by first identifying general information about the interviewee and the company they represent, e.g., number of employees or year of establishment. Following, we asked selected questions from a prepared interview guide for each interview based on the expertise of the interviewee and the course of the interview. Our interview guide is designed around questions that focus on exploring the alignment between a company's business processes and its ERP system. The complete interview guide is available online[1]. If appropriate, we asked individual follow-up questions to clarify the responses based on the interviewees' answers. The individual responses were collected, interpreted, and categorized by their meaning to identify the interviewees' needs using the Gioia Matrix (Gioia et al., 2013). Hence, we first identified 1st order concepts, deducted 2nd order themes, and finally identified aggregated dimensions that represent our presented challenges of current ERP systems.

## Observed Challenges of ERP Systems

By creating the Gioia Matrix (see Figure 1), we identified three primary challenges of current ERP systems. The following sections describe the captured challenges and their causes.

### Insufficient Transparency in Business Processes

The first challenge expresses that business processes involving ERP systems often lack sufficient transparency for end-users. We identified hard-coded, implicit business processes in ERP systems and lack of (up-to-date) process documentation in our interviews as reasons.

**Implicit business processes.** We learned that ERP transactions' control and data flow is implicitly represented by the successively generated documents required in business processes, e.g., sales orders, delivery notes, and invoices. In SAP S/4HANA, this sequence of documents is described as the document flow (SAP SE, 2021). However, none of our interviewees mentioned that implemented control and data flows of processes are graphically represented in ERP systems. Instead, the underlying business process is hidden in the respective implementation, limiting the comprehensibility of the ERP business processes, especially for non-technical users.

---

[1] Interview guide: https://doi.org/10.5281/zenodo.7853035

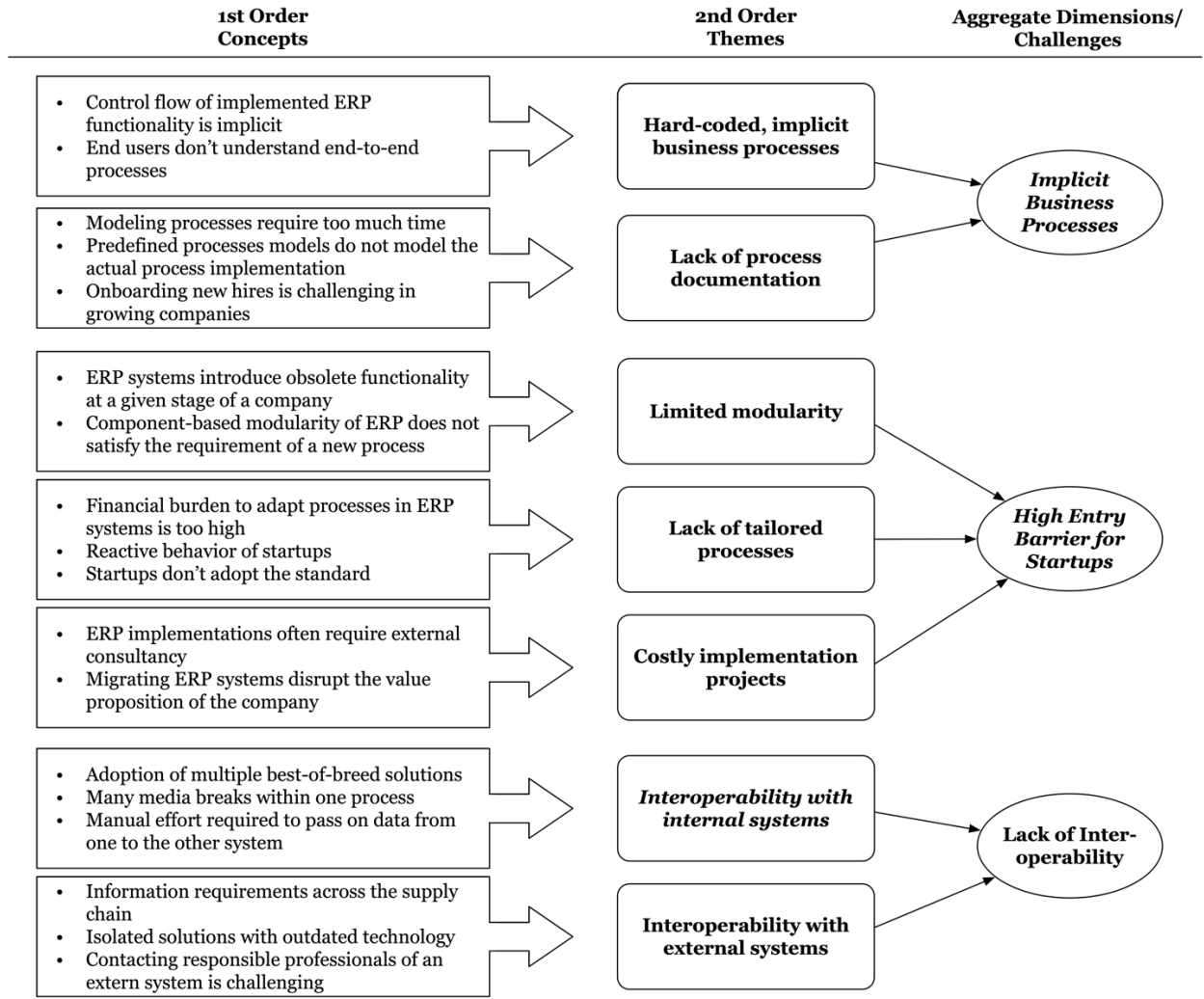

**Figure 1. Gioia Matrix: From interview observations to challenges**

**Lack of process documentation.** We observed that textual documentation of business processes often becomes outdated since they are only created during initial implementation or migration projects. When processes change, usually, the documentation is not adapted. Our interviewees cited the high manual effort as the reason, supporting existing literature (Ungan, 2006).

## High Entry Barrier for Startups

The introduction of an ERP system comes with a high entry barrier. According to the Federal Office of Statistics of Germany, only 31% of companies with 10 to 49 employees use ERP software, compared to 81% of larger companies with 250 and more employees (Destatis, 2021). We identified three main reasons contributing to a high entry barrier: limited modularity, the lack of tailored processes, and costly implementation projects.

**Limited modularity.** Each software component of an ERP system introduces further internal dependencies. State-of-the-art ERP systems, such as SAP S/4HANA, offer software components that represent sets of functions for different lines of business, like asset management or finance. The software components have dependencies, resulting in an increased configuration effort if new features are added. In addition, the components often include excessive functionality exceeding the requirements of the company's current state. For startups, a high degree of redundant functionality hinders the company from focusing on its value-adding business. The component dependencies also affect ERP systems' usability.

**Lack of tailored processes.** Another reason for the high entry barrier of ERP systems is the adoption of standard processes (Quiescenti et al., 2006). Most ERP systems impose a concrete process the company must comply with. Even if standard processes imply advantages such as cost minimization and improved process coordination, they provide only a limited opportunity to achieve competitive differentiation (Seddon, 2005). While this is sufficient for supporting processes, for example, human resources or finance, key processes that serve the company's unique goal usually need to be highly individual. Especially startups require tailored processes to emphasize their respective unique selling points.

**Costly implementation projects.** Introducing an ERP system has a reputation for being expensive and time-consuming (Schwenk, 2012). Most ERP systems implementations require a large upfront project to identify required configurations and data that must be migrated (Khanna & Arneja, 2012). While this phase allows the company to streamline its existing processes, it is costly since the required competence often comes from external consultancy (Dunaway, 2012) and poses the risk of disrupting daily business (Ahmad & Cuenca, 2013). The fear of disruption of the business and the costly upfront project were the top two reasons mentioned in the interviews that delayed the implementation or migration to a new ERP system.

## Lack of Interoperability

Another challenge of ERP systems is the lack of interoperability with systems within the same IT application landscape and external systems owned by business partners.

**Interoperability with internal systems.** Startups introduce different SaaS products to meet their business needs. This often happens in response to currently occurring problems. But even established, larger enterprises follow a best-of-breed IT strategy by implementing function-specific SaaS systems from selected strategic partners. However, an increasingly heterogeneous IT landscape consisting of systems in multiple cloud environments from multiple vendors often comes with integration challenges and redundant data storage. In our interviews, we observed that current business systems often do not support efficient interoperability with external applications, leading to high effort for data migration and integration of processes. One interviewee shared that one reason for excluding ERP systems in the selection phase of an implementation project was the inability to integrate with other existing systems in the company. That highlights the importance of integration capabilities for business applications in accordance with literature researching the selection of ERP systems (Wei et al., 2005).

**Interoperability with external systems.** Today's end-to-end processes span entire value chains and go beyond company boundaries, making integration possibilities with external systems increasingly important. The fact that current ERP systems only focus on one enterprise instead of supporting the complete value chain of the business increases this problem. The interoperability between systems of other enterprises is rarely supported by the ERP system (Lyytinen & Damsgaard, 2011). The development of interfaces between these ERP systems is often associated with considerable communication overhead.

# A Vision for Process-centric Business Process Platforms

Current ERP systems face several challenges of both technical and organizational nature. We believe that many of the current technical issues of ERP systems occur due to ERP systems' architecture. Hence, we propose a novel class of enterprise systems called Business Process Platforms (BPP). This section illustrates a vision for process-centric BPPs. We present three enablers BPPs should utilize to achieve a better TTF for their implementing organizations. Figure 2 shows how the enablers, based on available supporting technologies, relate to the challenges.

## Business Processes as First-class Entities

We argue that it is time to consider business processes - besides data - as first-class entities to design BPPs that address the challenges of insufficient transparency and costly process changes. To achieve process centricity, we believe that a BPP should be built on modular, adaptable, and graphically represented executable business processes.

**Executable business processes.** A BPP should utilize executable business processes comparable to existing workflow engines such as Camunda Platform or Netflix Conductor. Hence, it must support the core capabilities of workflow engines like persistence, scheduling, and versioning. (Rücker, 2021).

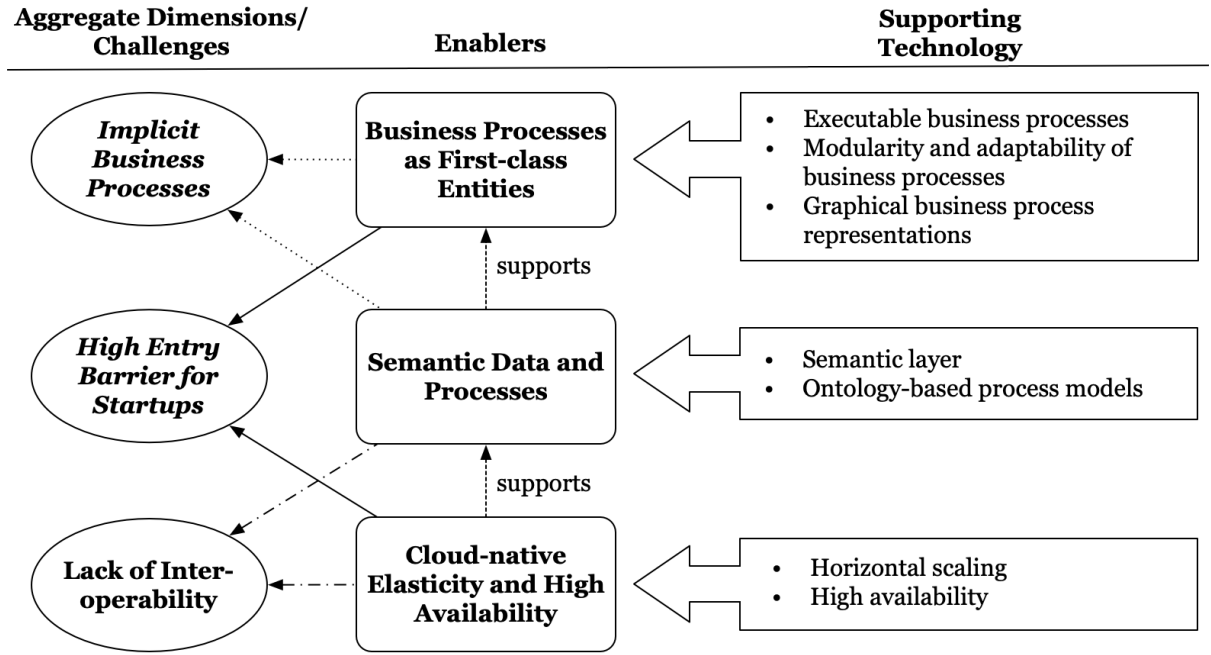

**Figure 2. Challenges of ERP systems and how BPPs address them**

**Modularity and adaptability of business processes.** Based on insights from our interviews, we argue that companies search for lightweight solutions to fulfill their business goals. In this regard, we infer that encapsulated business processes are superior to traditional ERP modules built for different departments, such as finance. While these modules provide extensive functionality, they usually depict only a subset of an end-to-end business process. To increase the modularity of future business applications, they must support extensions via independently executable business processes instead of traditional ERP components. Having modular business processes instead of ERP components enables a company to add and remove processes without the burden of exceeding functionality. In addition, the platform must provide pre-defined processes with varying characteristics via templates to cope with changing requirements of business processes. Each business process should exist in multiple versions to reflect stages of growth and industries to enable users to upgrade individual processes. Additionally, each process must enable adjustments to customize pre-defined templates to a company's requirements. Three user groups create such process customizations based on a template. First, logical changes, such as adjustments to the control flow of a process, can be implemented by business users. Second, adjustments requiring additional, not yet supported functionality should be implemented by developers via an abstract interface and integrated as a process task according to the semantic data schema. Finally, compliance-related changes, such as temporary tax subsidies, are reflected in the semantic data model and updated by the BPP's vendor. A company is thus able to upgrade distinct processes by using an equivalent process template designed for larger company sizes when the enterprise outgrows the existing process. If a company requires adaptations not covered by the provided process templates, each process must be adjustable to enable the end-user to customize the processes based on their needs. The end-user can adjust the control flow of the process using a visual editor, whereas more extensive adjustments requiring new process tasks are individually programmed. Such tasks are designed using abstract interfaces of the BPP instead of concrete solutions so that external software can be integrated as a process task. The BPP's provider must specify these abstract interfaces. Process adjustments by the company can be captured and moved forward in case of a process upgrade without requiring a large migration project because underlying services remain stable, and only processes that call these services are adapted. Adapted business processes could be offered as new templates for other companies.

**Graphical business process representations.** End-users of ERP systems need process transparency. Therefore, we argue that the User Interface of future ERP systems should be based on business process models. The process models should represent the implemented business processes in a human-understandable notation such as BPMN. The process models should enable all process stakeholders to better understand the business process, the available functionality, and the respective data model. The visual representation can be extended through process-based KPIs, such as time per task and cycle time, directly embedded into the process model to give relevant feedback on the process performance to the user. Currently, process mining is used to derive the actually executed process models based on historical data (van der Aalst et al., 2012). With a BPP, we envision that business software is fundamentally based on such process models, making them executable and graphical from the beginning.

## Semantic Data and Processes

Current business software does not support detailed descriptions of the stored data, which would allow machines or humans an in-depth understanding of its underlying meaning. We argue that a BPP should use semantically enriched data and processes to address the interoperability challenge of current ERP systems. Following the example of the semantic web, we envision the data schemas used in business processes to be enriched with standardized descriptions that simplify data integration. This involves an additional semantic layer and executable ontology-based process models on top of the data layer.

**Semantic layer.** Usually, ERP systems store their data in relational databases. However, their data is often not expressive enough to meet the high interoperability demands of today's heterogeneous supply chains and IT system landscapes. Since end-to-end processes can go beyond the borders of single systems and even companies (Bender et al., 2021), it must also be possible to execute cross-system processes. Using ontologies, for instance, enables true semantic integration of different systems' data across businesses (Gardner, 2005). Hence, introducing an additional semantic layer on top of the data model improves data integration capabilities and thus reduces the manual effort for end-users. Furthermore, the semantic integration of data offers the advantage of making data discoverable with established semantic web technologies, such as the Resource Description Framework and the SPARQL language.

**Ontology-based process models.** Semantically enriched data can build the foundation to enable the execution of process models because they provide understandable information for both humans and machines (Corea et al., 2021). The mapping of business entities from ontologies to traditional process models, resulting in ontology-based process models, bridges the gap between processes' data and control flow. Also, ontology-based process models provide BPPs with advanced ontology-based analytics, such as the capability to query process information using SPARQL, leading to reduced risks and costs of process changes (Corea et al., 2021). Possible use cases include the detection of compliance issues in process models (Corea & Delfmann, 2017), the creation of various process views that hold information, such as organizational hierarchies, or how process resources interact (Adams et al., 2021).

## Cloud-native Elasticity and High Availability

To use the full potential of cloud infrastructure, we argue that a future BPP has to embrace cloud-native technology. Cloud-native infrastructure offers elasticity and fault tolerance for BPPs, building the technical foundation for the other enablers. Furthermore, it addresses the challenges of a high entry barrier for startups and the lack of interoperability by leveraging the adoption of respective workloads for a given situation through horizontal scaling and high availability offered by the cloud.

**Horizontal scaling.** A company's IT infrastructure must scale with the requirements and load an IT system is experiencing while a company grows. Companies might experience a massive increase in transactions in a short period or a steady increase in workload due to the business' growth. In both cases, the systems have to deal with the higher throughput without impacting the current business, independent of the time frame in which the demand rises. At the same time, the system should be able to scale down if required to save costs. Horizontal scaling capabilities are a well-known way to address the scalability demands of business applications. Most of today's ERP systems are built as monolithic applications, not allowing selective scaling of individual functions of the system. Monolithic systems do not leverage the potential of the cloud due to missing architectural adjustments of the application to the cloud environment (Balalaie et al., 2016).

**High availability.** Companies cannot tolerate downtimes from planned maintenance or system failures. Whereas resilience and fault tolerance are favorable future-proof qualities for SMEs, it is crucial for large enterprises to have their system available all the time. One interviewee said that it is challenging to keep the upgrade process of the existing ERP system within the boundaries of one weekend. They also mentioned that relocating critical parts of their infrastructure to other data centers without impacting the ongoing business reduces business interruptions. Infrastructural resilience and fault tolerance are impossible without the proper systems architecture.

## Discussion

Effective use of technology depends on its alignment with the specific tasks that users need to perform, as described by the TTF theory (Goodhue & Thompson, 1995). Goodhue and Thompson describe eight main factors contributing to task-technology fit: quality, locatability, authorization, compatibility, ease of use, production timeliness, systems reliability, and the relationship with users. Our study revealed several challenges connected to these factors that growing companies face when employing ERP systems, pointing to a mismatch between tasks and technology in this domain.

We identified that the studied companies lacked a consolidated, transparent view of their business processes. This finding is closely related to the TTF factors of *quality* and *locatability*, describing the ability to clearly identify relevant data sources and to understand the meaning of data points on a technical and business level in adequate detail. Without comprehensive insights into the executed process, locating the set of relevant data of sufficient quality needed as process inputs is challenging. Our vision addresses these challenges by visually representing executable business processes enriched with semantic data. While visual representations benefit understandability, attention must be paid to ensure human comprehension of the created models (Figl, 2017), which is affected by their size (Dikici et al., 2018). Therefore, business processes must be hierarchically decomposed and represented on different abstraction layers to limit their size and complexity. The envisioned modularity through processes lowers the barrier for startups to introduce ERP systems, as only the currently required processes need to be analyzed and implemented. We envision process templates provided by third parties with deep domain knowledge that help with this initial adoption. By enabling every process to be individually upgradable using a template of a larger company in the same industry, we address the TTF factor of *ease of use*. Dynamically adapting the amount and complexity of offered ERP-supported processes over time prevents harsh breaks in usability and enables continuous user training. Making use of modern cloud services allows a focus on application design (as opposed to IT infrastructure management) and provides for the TTF factor of *systems reliability* (i.e., the system "uptime") at an industry-standard level. However, security, privacy, and connectivity risks of the employed cloud platforms must be considered (Avram, 2014). Additionally, the TTF factor of *compatibility* is addressed as part of our vision. It refers to the ability to effectively combine data from disparate sources, which requires a thorough understanding and overview of the available data within organizations. We propose using a semantic data model to improve interoperability and system compatibility, which requires different applications to agree on shared data definitions (Wegner, 1996).

While our explorative study design with industry experts allowed first-hand insights into the prevailing problems in practice, it also comes with limitations. First, our findings are limited to the 15 interviews we conducted and the questions we asked. However, we reached a saturation point of insights by getting repeating answers in multiple subsequent interviews. The selection of interviewees was mainly based on existing industry contacts. Other interviewees with different backgrounds could have provided different insights, affecting current ERP systems' highlighted challenges. Second, this paper only focuses on the challenges the interviewees explicitly and repeatedly mentioned. Topics such as data security, safety, or IT governance, which are also very important in an enterprise, still need to be considered. Although the challenges of current ERP systems inspire our vision, we acknowledge that our solution may not be equally suitable for companies of all sizes. Its modularity and flexibility make it particularly well-suited for startups and growing businesses that need to adapt quickly to changing circumstances. However, larger enterprises that have already established stable operations may not benefit as much from our solution, as they are less likely to require frequent system modifications. Furthermore, our vision is based on our interpretation of how to solve the derived challenges. Other researchers and practitioners might come to different conclusions. Finally, there are still many open technical questions regarding the concrete implementation of our vision. These are subjects of future research and other conceptual questions, such as how adjustments to the process model are reflected in task implementations and how process development environments should be designed.

## Conclusion

The history of enterprise systems has resulted in an accumulation of different architectural decisions. However, some of these decisions no longer align with today's highly decoupled software service landscape and the need for flexible processes, leading to technical challenges. In this paper, we shared first-hand insights into the challenges of ERP systems based on 15 interviews using purposeful sampling. We applied the Gioia Matrix methodology to identify practitioners' challenges in growing companies. Based on the interviews, the main challenges are a high entry barrier for startups, insufficient transparency in business processes, and a lack of interoperability. We propose a vision of future ERP systems called BPP to address the challenges. A BPP defines itself through well-defined, adjustable business processes which can grow with the company. To overcome the identified challenges of current ERP systems, the foundation of the envisioned BPP are three technical enablers: business processes as first-class entities, semantic data and processes and cloud-native elasticity, and high availability. Our vision for a new generation of ERP systems addresses the practical challenges of current ERP systems. Future research might investigate how BPPs are used and how they impact companies' operations. In a design science research paradigm, a BPP prototype could be implemented and evaluated in real-life settings.